# Pascal Program for Generating Tables of $SU(3)$ Clebsch-Gordan Coefficients [*]

Thomas A. Kaeding

*Theoretical Physics Group*
*Lawrence Berkeley Laboratory*
*University of California*
*Berkeley, California 94720*

**Abstract**

Pascal routines are provided that generate representations of the group $SU(3)$ and tabulate the Clebsch-Gordan coefficients in the products of representations.

[*]This work was supported by the Director, Office of Energy Research, Office of High Energy and Nuclear Physics, Division of High Energy Physics of the U.S. Department of Energy under Contract DE-AC03-76SF00098.

# Disclaimer

# PROGRAM SUMMARY

*Title of Program:* SU3Clebsch

*Catalogue number:*

*Program obtainable from:* CPC Program Library, Queen's University of Belfast, N. Ireland (see application form in this issue)

*Licensing provisions:* Persons requesting the program must sign the standard CPC non-profit use licence (see licence agreement printed in every issue).

*Computer for which the program is designed and others on which it has been tested:* VAX workstation 4090, VAX 6610

*Operating systems under which the program has been tested:* VMS v. T6.1

*Programming language used:* Pascal (DEC Pascal v. 4.4)

*Memory required to execute:* Variable; 4 M words for $\bar{\mathbf{3}} \otimes \mathbf{3}$, 10.8 M words for $\bar{\mathbf{10}} \otimes \mathbf{8}$.

*No. of bits in a word:* 32

*No. of processors used:* 1

*Has the code been vectorized?* No

*No. of lines in distributed program:* 8250

*Classification:* 4.2 Computational methods of algebras and groups, 11.6 Phenomenological and empirical models and theories in elementary particle physics.

*Nature of physical problem:*

Calculations in models based on the group $SU(3)$ often require Clebsch-Gordan coefficients in the product of group representations. Previously compiled tables [1] contained only representations having integral hypercharge values. They are therefore useful for calculations involving mesons and hadrons, but are inadequate for calculations that involve arbitrary numbers of quarks.

*Method of solution:*

The program generates representations of $SU(3)$ whose states are represented by vectors. The dimension of such a vector is the number of $SU(3)$ tensor indices used to describe the representation. The vectors in the product of two representations are formed by taking outer products. The set of such vectors is decomposed into the irreducible representations of the Clebsch-Gordan series. The Clebsch-Gordan coefficients are then found as inner products between the initial set of vectors and the states in this decomposition.

*Restrictions on the complexity of the problem:*

The program is written to handle representations that have at most nine $SU(3)$ indices. For most physical calculations this should suffice.

*Typical running time:*

Running time depends on which table is compiled. Times quoted are for the VAX 4090. For $\bar{\mathbf{3}} \otimes \mathbf{3}$, the CPU time is 4.3 sec. For $\bar{\mathbf{10}} \otimes \mathbf{8}$, the CPU time is 52 min 26 sec.

*Unusual features of the program:*

The program does not compile a general table of coefficients. However, the calculation of many tables of coefficients has been written into the program. If the user needs any other tables, he is encouraged to follow the examples in the code in order to write other routines.

# LONG WRITE-UP

## 1 Introduction

There exist in the literature tables of Clebsch-Gordan tables involving representations of SU(3) with integral hypercharge values [1]. These are inadequate for handling physical problems that use representations that cannot be built from combinations involving three quarks or quark plus antiquark. Such problems include calculations involving exotic hadrons, sextet quarks, or flavor $SU(3)$ transitions such as $c \to s$. Hence the need to develop the program SU3Clebsch that calculates tables involving representations with nonintegral hypercharge.

The conventions of [2] are used to label the representations. Users desiring a review of the theoretical background can consult [3].

## 2 Calculational Method

We can always write an $SU(3)$ tensor with only upper indices which run over $u$=1, $d$=2, $s$=3. In the program tensors are represented by arrays of integers. Real values are not necessary as a result of the structure of the group; i.e., the quantum numbers of states are not continuous and therefore the coefficient of each term in a vector is a rational number. The number of integers is $3^n$ for an array that represents a tensor with $n$ indices. A state in a representation of $SU(3)$ is a tensor with a particular hypercharge, isospin and third component of isospin. The array representing such a state is called a "vector" in the program code. States of a representation must be properly normalized in order to maintain the unitarity of the table of Clebsch-Gordan tables. That is, the inner product of a vector with itself must be 1. Therefore the sum of the squares of the integers in a vector is stored in the 0, 1, ..., 1 component of the vector and serves as the denominator (squared). The integers themselves serve as the numerators. The vectors can now be thought of as arrays of rational numbers. For example, consider the highest-weight state of the **8**. In the pion octet this state is the $K^+$, and in the lowest baryon octet it is the proton.

$$\begin{aligned} \mathbf{8}\left(y=1, i=\tfrac{1}{2}, i_3=\tfrac{1}{2}\right) &= A^{ijk}, \\ A^{uud} &= \tfrac{1}{\sqrt{2}}, \\ A^{uud} &= -\tfrac{1}{\sqrt{2}}. \end{aligned} \quad (1)$$

The vector that represents it has nonzero components only in the spaces marked by $u$, $u$, $d$ and $u$, $d$, $u$. The norm squared is 2 and is stored in the space marked $0$, $u$, $u$. We can write it simply as

$$\mathbf{8}\left(y=1, i=\frac{1}{2}, i_3=\frac{1}{2}\right) = \frac{1}{\sqrt{2}}(uud - udu). \quad (2)$$

Although our storage method is not efficient, it has facilitated the coding of the program.

The raising and lowering operations of the group $SU(3)$ are performed by subroutines that raise or lower the indices of a vector. Raising or lowering can occur in any of the three directions corresponding to the three $SU(2)$ subgroups of SU(3). For example, if we wish to lower the isospin of the highest-weight state of the **8** shown above, the first nonzero component can have either its first or second index lowered, while the other nonzero component can have either the first or third index lowered. Two of these possibilities cancel to leave

$$\mathbf{8}\left(y=1, i=\frac{1}{2}, i_3=-\frac{1}{2}\right) = \frac{1}{\sqrt{2}}(dud - ddu). \quad (3)$$

The raising and lowering operations are used to construct all of the vectors in a given representation from the highest-weight state.

The product of two representations is calculated as all possible outer products of vectors in the representations. The outer product of an $n$-vector and an $m$-vector is an $(n+m)$-vector whose nonzero components are the products of the appropriate nonzero components of the factor vectors. For example, the outer product of the highest-weight vector of the **8** in equation (2) and the highest-weight vector of the $\bar{\mathbf{3}}$,

$$\bar{\mathbf{3}}\left(y=\frac{2}{3}, i=0, i_3=0\right) = \frac{1}{\sqrt{2}}(ud - du), \quad (4)$$

is

$$\begin{aligned} \mathbf{8}\left(y=1, i=\tfrac{1}{2}, i_3=\tfrac{1}{2}\right) &\otimes \bar{\mathbf{3}}\left(y=\tfrac{2}{3}, i=0, i_3=0\right) \\ &= \tfrac{1}{2}(uudud - uuddu - uduud + ududu). \end{aligned} \quad (5)$$

The representations in the Clebsch-Gordan series are constructed from the largest to the smallest. By larger we mean having a highest-weight vector with quantum numbers larger than the highest-weight vector of a smaller representation. The smaller ones are constructed such that their highest-weight vectors (and hence all of their vectors) are orthogonal to the vectors of the larger representations of the same weight. For example, consider the product (5) above. This is the highest-weight vector of the $\bar{15}$ representation. We apply a lowering operation to find the state of the $\bar{15}$ that has the set of quantum numbers ("weights") $y = \frac{2}{3}$, $i = 1$, $i_3 = 1$, i.e.,

$$\bar{15}\left(y = \tfrac{2}{3}, i = 1, i_3 = 1\right)$$
$$= \tfrac{1}{\sqrt{8}}(uudus + uusud - uudsu - uusdu - uduus - usuud + udusu + usudu). \tag{6}$$

The highest-weight vector of the **6** in the product is found to be a vector spanned by the outer products of the vectors in **8** and $\bar{3}$ of the same weight and which is orthogonal to this vector of the $\bar{15}$. The complete $\bar{15}$ and **6** representations are constructed, and the $\bar{3}$ in the product is constructed to be orthogonal to both.

To find the Clebsch-Gordan coefficients, we consider two sets of vectors. The first set consists of all possible outer products of the vectors in the representations that we are multiplying. The second set consists of all of the vectors in the representations that have been constructed in the product. The coefficients are the inner products between these sets of vectors. In the example that we have been following, the coefficient

$$< \bar{15}\left(y = \tfrac{2}{3}, i = 1, i_3 = 1\right) | \mathbf{8}\left(y = 1, i = \tfrac{1}{2}, i_3 = \tfrac{1}{2}\right) \bar{\mathbf{3}}\left(y = -\tfrac{1}{3}, i = \tfrac{1}{2}, i_3 = \tfrac{1}{2}\right) >$$
$$= \tfrac{1}{\sqrt{8}}(uudus + uusud - uudsu - uusdu - uduus - usuud + udusu + usudu)$$
$$\cdot \left(\tfrac{1}{\sqrt{2}}(uud - udu) \otimes \tfrac{1}{\sqrt{2}}(us - su)\right)$$
$$= \tfrac{1}{\sqrt{2}}. \tag{7}$$

This is the second line in the table of sample output shown below.

## 3 Program Structure

The program is broken up into routines that handle vectors and representations with specific numbers of $SU(3)$ indices. This means that we have replication of procedures for each set of indices, but that the program is easier to use and to code. The main routine calls the subroutines that generate the tables desired. They in turn call the routines that manipulate the vectors and representations necessary to compile the tables.

The routines that build the pieces necessary to construct the tables of coefficients are named **doRxS**, where **R** and **S** are the names of the representations to be multiplied. The user must remove the comment braces around the procedure calls in the main routine (at the bottom of the program listing) for the tables that he wishes to calculate. The routines that are included in this version of the program are

| | | | | | | | |
|---|---|---|---|---|---|---|---|
| do3barx3    | $\bar{3}$  | $\otimes$ | $3$        | do3barx3bar | $\bar{3}$  | $\otimes$ | $\bar{3}$ |
| do6x3       | $6$        | $\otimes$ | $3$        | do6x3bar    | $6$        | $\otimes$ | $\bar{3}$ |
| do6x6       | $6$        | $\otimes$ | $6$        | do6barx3    | $\bar{6}$  | $\otimes$ | $3$       |
| do6barx3bar | $\bar{6}$  | $\otimes$ | $\bar{3}$  | do6barx6    | $\bar{6}$  | $\otimes$ | $6$       |
| do6barx6bar | $\bar{6}$  | $\otimes$ | $\bar{6}$  | do8x3       | $8$        | $\otimes$ | $3$       |
| do8x3bar    | $8$        | $\otimes$ | $\bar{3}$  | do8x6       | $8$        | $\otimes$ | $6$       |
| do8x6bar    | $8$        | $\otimes$ | $\bar{6}$  | do8x8       | $8$        | $\otimes$ | $8$       |
| do10x3bar   | $10$       | $\otimes$ | $\bar{3}$  | do10x8      | $10$       | $\otimes$ | $8$       |
| do10barx8   | $\bar{10}$ | $\otimes$ | $8$        | do15x3bar   | $15$       | $\otimes$ | $\bar{3}$ |
| do27x3bar   | $27$       | $\otimes$ | $\bar{3}$  | | | | |

The functions that the above routines use are

| | |
|---|---|
| `index`         | Returns $u$, $d$, $s$ as the $SU(3)$ index for 1, 2, 3 |
| `nullvector`$n$ | Returns true if its $n$-vector argument is null |
| `inner`$n$      | Finds the inner product of two $n$-vectors |
| `divisible`$n$  | Returns true if its $n$-vector argument is divisible by its integer argument |
| `orthogonal`$n$ | Returns true if its two $n$-vector arguments are orthogonal |
| `normalvertical` | Returns true if all columns in the table of coefficients have unit norm |
| `normalhorizontal` | Returns true if all rows in the table of coefficients have unit norm |

The procedures used are

| | |
|---|---|
| `initialize`$n$ | Initializes an $n$-vector |
| `initrep`$n$ | Initializes a rep composed of $n$-vectors |
| `inittable` | Initializes the table of coefficients |
| `reducefraction` | Reduces a fraction to its lowest form |
| `scalardivide`$n$ | Divides an $n$-vector by an integer |
| `normalize`$n$ | Finds the square of the norm of an $n$-vector and stores it in the vector's 0th component |
| `scalarmultiply`$n$ | Multiplies an $n$-vector by an integer |
| `addvectors`$n$ | Adds two $n$-vectors |
| `T_`$n$ | Performs the $SU(3)$ raising and lowering operations on $n$-vectors |
| `highestw`$n$ | Determines whether its $n$-vector argument is a highest-weight vector |
| `findperp`$n$ | Finds the projection of one $n$-vector onto the subspace orthogonal to another $n$-vector |
| `calcR_`$n$ | Calculates the representation **R** from its highest-weight vector in terms of $n$-vectors; only the **R** and $n$ combinations needed by the procedures `doRxS` are included |
| `writerep`$n$ | Writes the $n$-vectors of a representation into `repfile` |
| `outer`$n$`x`$m$ | Finds the outer product of an $n$-vector and an $m$-vector |
| `tabulate`$n$`x`$m$ | Tabulates the coefficients into `table` and into the output files for the product of a representation of $n$-vectors and of $m$-vectors to a representation of $(n+m)$-vectors |
| display**R** | Writes the weight diagram for the representation **R** into `repfile` |
| checktable | Calls `horizontalnormal` and `verticalnormal` and checks that `table` has the correct number of columns filled |

The global variables used by the program are

| | |
|---|---|
| `outfile` | File into which the table of coefficients is written |
| `outlog` | Log file for error messages and a duplicate set of coefficients |
| `repfile` | File into which the vectors comprising a representation are written |
| `name` | String which is the name of `repfile` |
| `isit` | A boolean variable used in highestw$n$ |
| `table` | The table of Clebsch-Gordan coefficients |
| `tablemarker` | Number of columns of `table` that have been filled |
| `size1`, `size2` | The sizes of the factor representations |

# 4  Using the Program

In order to use this program, the user must remove the comment braces around the call to **doRxS** in the the main procedure for the table that he wishes to calculate. Then the code can be compiled and run. We recommend calculating one table at a time, due to the variable, and sometimes very long, running time.

Since we store representations as arrays of vectors, the vectors are labelled not by their quantum numbers, but merely by integers. Here we explain the ordering of the vectors. The highest-weight state is always numbered 1. This is the state of highest isospin in the isomultiplet of highest hypercharge. States are then sequentially numbered from right to left and then from top to bottom of the weight diagram of the representation. For representations that have doubly (or more) occupied sites in the weight diagram, the states with largest total isospin are numbered first. Only after one state in every site has been numbered are the other states numbered, again from right to left and from top to bottom. The procedures **displayR** write the weight diagrams with their numbering schemes into the files `repfile`. The figure shows the fifteen representation and the numbering of its states. It contains the isoquartet 4, 5, 6, 7, the isotriplet 8, 9, 10, the isodoublet 13, 14, and the isosinglet 15.

A square root is assumed to appear over each coefficient in the tables generated. Any minus signs are outside the square root.

If the user finds it necessary to write additional procedures to compile additional tables, then he should be aware of some error messages that may be written to `outlog`. The routines `findperp`$n$ write error messages if their results are not highest-isospin vectors or if their results are null vectors. In many cases, one does not want the highest-isospin vector, and so this error message is often ignored. When **calcR_**$n$ can ignore that message, it will place a note to that effect in `outlog`. If `findperp`$n$ returns a message that the result is a null vector, it is an indication that the desired result was not found (because the two input vectors were parallel). A new choice of input vectors is then required. If a routine returns an error message saying that a vector is not a highest-weight vector when one was expected, then this is an indication that the construction of that vector was flawed. This can only be corrected on a case-by-case basis.

## 5  Sample Output

Below is a sample table generated by the program. The first column is the number of the state in the first representation in the product. The second column is the number of the state in the second representation. The third column is the number of the state in the product representation that is denoted in the heading of that part of the table. The last two columns are the numerator and denominator of the Clebsch-Gordan coefficient between those three states. Recall that a square root is assumed to appear over each fraction. Only nonzero coefficients are tabulated.

## 6  Acknowledgement

This work was supported by the Director, Office of Energy Research, Office of High Energy and Nuclear Physics, Division of High Energy Physics of the U.S. Department of Energy under Contract DE-AC03-76SF00098.

## Figure Captions

Figure 1:    Labelling of the states of the **15**.